\documentclass[prb,showpacs,byrevtex,twocolumn]{revtex4-1}
\usepackage{graphicx}
\usepackage{amsmath}
\usepackage[section]{placeins}
\usepackage{appendix}
\usepackage{color}

\begin{document}

\title{The Heisenberg spin-1/2 XXZ chain in the presence of electric and magnetic fields}

\author{Pradeep Thakur and P. Durganandini}

\affiliation{Department of Physics, Savitribai Phule Pune University, Pune-411007, INDIA}

\date{\today}

\begin{abstract}
We study the interplay of electric and magnetic order in the one dimensional Heisenberg spin-1/2 XXZ chain with large Ising anisotropy in the presence of the Dzyaloshinskii-Moriya (D-M) interaction and with longitudinal and transverse magnetic fields, interpreting the D-M interaction as a coupling between the local electric polarization and an external electric field. We obtain the ground state phase diagram using the density matrix renormalization group method and compute various ground state quantities like the magnetization, staggered magnetization, electric polarization and spin correlation functions, etc. In the presence of both longitudinal and transverse magnetic fields, there are three different phases corresponding to a gapped N\'{e}el phase with antiferromagnetic (AF) order, gapped saturated phase and a critical incommensurate gapless phase. The external electric field modifies the phase boundaries but does not lead to any new phases. Both external magnetic fields and electric fields can be used to tune between the phases. We also show that the transverse magnetic field induces a vector chiral order in the N\'{e}el phase(even in the absence of an electric field) which can be interpreted as an electric polarization in a direction parallel to the AF order. 
\end{abstract}


\maketitle

\section{Introduction}
The one dimensional Heisenberg spin $S= 1/2$ XXZ model has long served as a paradigm for the study of quantum magnetism~\cite{mikeska-review} in low dimensions. The study of effects induced by external magnetic fields has been of particular interest since the magnetic behaviour in an external magnetic field can be qualitatively different depending on the magnitude and direction of the external field~\citep{yang1, *yang2, *yang3, kurmann-thomas-muller, mori, hiki-furu-1998, hieida, dko2002, dkol, dmitriev-Ising, caux, dmitri-krivnov, hiki-furu-2004, hagemans}. This has led to interest in the study of field-induced quantum phase transitions~\cite{mikeska-review, yang1, *yang2, *yang3, kurmann-thomas-muller, dko2002, dkol, dmitriev-Ising, caux, dmitri-krivnov, hagemans, giamarchi-schulz, chitra-giamarchi, sato-oshikawa}. Such models are also believed to describe many quasi-one dimensional compounds like $\text{TlCoCl}_3$ ~\cite{oosawa},$\text{CsCoCl}_3$~\cite{shiba-ueda},$\text{Cs}_2\text{CoCl}_4$~\cite{kenzelmann,breunig2015},$\text{SrCo}_2\text{V}_2\text{O}_8 $~\cite{bera2015, *bera}. In recent years, there has also been a lot of interest in the study of the interplay of electric and magnetic order in quantum spin systems, motivated by the direct correlation between ferroelectricity and non-collinear magnetic order discovered in perovskite multiferroics like $\text{RMnO}_3$ with $\text{R}= \text{Tb, Dy, Gd}$ and $\text{Eu}_{1-x}\text{Y}_x$ and in several edge-sharing copper oxide compounds like $\text{LiCu}_2\text{O}_2$, $\text{LiCuVO}_4$, etc~\cite{mf-review}. The direct correlation between magnetic order and ferrolectricity implies the existence of an intrinsic magnetoelectric effect in these systems. This has led to a resurgence of focus in the study of the magnetoelectric effect due to the interest in being able to manipulate magnetism with electric fields and vice versa. A microscopic model which connects the (ferro-)electric polarization with the non-collinear magnetic ordering of neighbouring spins is a lattice geometry independent model based on the spin current mechanism~\cite{knb}:
\begin{equation}
\vec{P_i} \sim \gamma \vec{e}_{ij} \times (\vec{S}_i \times \vec{S}_j)
\end{equation}
\noindent where $\vec{P}_i$ is the local polarization at the $i^{th}$ site, $\vec{e}_{ij}$ is the unit vector pointing from site $\text{i}$ to site $\text{i+1}$, $S_i^{a}$ ($a=x,y,z$) is the $a^{th}$ component of the spin operator at the $i^{th}$ site and $\gamma$ is a material-dependent constant. The electric polarization operator is also intimately related to the Dzyaloshinskii-Moriya (D-M) interaction~\cite{DM1, *DM2},$\vec D_{ij}\cdotp \vec S_i \times \vec S_ j͒$, which arises due to spin-orbit coupling. 
The spin-orbit interaction leads to canting of the spins and hence frustration in the spin system which makes the study of spin orbit effects in spin chain systems of special interest. Besides, the D-M interaction is known to modify the dynamic properties  and quantum entanglement of spin chains~\cite{derzhkoetal,jafarietal}. It also plays an important role in explaining the electron spin resonance experiments on one-dimensional antiferromagnets~\cite{oshikawa-affleck1, *oshikawa-affleck2}.   

Recent studies of the magnetoelectric effect in the anisotropic XXZ spin chain with large anisotropy in the presence of magnetic field and the D-M interaction with the latter being interpreted as an external electric field ~\cite{brockmann, pradeep} showed that the electric field can be used to tune between phases. Different regimes of electric polarization with the different regimes being controlled by the magnetic field were obtained. For a pure transverse magnetic field, there are two different regimes of polarization corresponding to the N\'{e}el phase and the fully magnetically saturated phase~\cite{pradeep}. For the case of a pure longitudinal magnetic field, there are three different regimes for the polarization corresponding to the N\'{e}el phase, the Luttinger liquid phase and the fully magnetically saturated phase~\cite{brockmann, pradeep}. An interesting question is that of the effect of the electric field when the magnetic field has both longitudinal and transverse components and the spin rotational symmetry is completely broken. For example, recent bosonization studies on the nearly isotropic spin-1/2 Heisenberg chain with longitudinal and transverse magnetic fields have shown that the competition between D-M interaction and longitudinal and transverse magnetic fields can lead to interesting field-induced antiferromagnetic order~\cite{ganga, garate-affleck}. The question is also of relevance in the context of neutron scattering experiments when the magnetic field is applied at an angle to the anisotropy axis. In this work, we address this question by studying the interplay of electric and magnetic order in the anisotropic spin-1/2 XXZ chain with large anisotropy with longitudinal and transverse magnetic fields and with the D-M interaction interpreting it as an electric field. We have numerically determined the ground state phase diagram using the density matrix renormalization group (DMRG) method. Various ground state quantities like the energy gap, magnetization, staggered magnetization, electric polarization and spin correlation functions have also been computed. Our main result is that for large Ising anisotropy, the electric field does not lead to any new phase but modifies the phase boundaries. With increasing electric field, the AF N\'{e}el phase region reduces while the incommensurate phase region grows. We also obtain the electric polarization induced by the external electric field and show that both transverse and longitudinal magnetic fields can be used to tune the electric polarization.  Interestingly, we show that the transverse magnetic field induces an electric polarization parallel to the AF order as well as nematic order even in the absence of the electric field. 

The plan of our paper is as follows: In Sec.~\ref{sec:XXZ}, we briefly discuss the $XXZ$ model in the presence of the longitudinal and transverse magnetic fields and present the results here for the phase diagram obtained using the DMRG study and compare with previous studies which were mainly based on the mean field approximation. There are some earlier numerical studies on the XXZ model in the presence of a longitudinal field~\cite{hiki-furu-2004} and a transverse field~\cite{caux} but as far as we are aware, there do not seem to be earlier detailed numerical studies of the phase diagram for the model in the presence of both transverse and longitudinal magnetic fields. We also show that the transverse magnetic field leads to nematic order as well as vector chirality in a direction parallel to the AF order in the N\'{e}el phase. In Sec.~\ref{sec:electric}, we present the results for the phase diagram obtained in the presence of the electric field when both the longitudinal and transverse fields are present. We discuss the behaviour of various ground state  quantities and their dependences on the external fields. We summarize our work in Sec.~\ref{sec:Conc}.

\section{\label{sec:XXZ} XXZ chain with longitudinal and transverse magnetic fields}

The anisotropic $S=1/2$ XXZ-chain in the presence of longitudinal and transverse magnetic fields and with
a  D-M interaction along the $z$ direction is described by the Hamiltonian $\mathcal{H}$: 
\begin{eqnarray}
\mathcal{H} = {\sum_{n=1}^N} [J(S_n^x S_{n+1}^x + S_n^y S_{n+1}^y + \Delta S_n^z S_{n+1}^z) \nonumber \\
+ E (S_n^x S_{n+1}^y - S_n^y S_{n+1}^x) - h_z S_n^z - h_x S_n^x] \label{eq:ham_dm}
\end{eqnarray}
where $S_i^a$, with $a = x, y, z$, denote the components of the spin operator at the $i$th site along the chain, $\Delta$ is the easy-axis anisotropy (the $xy$-plane being the easy-plane), $ h_x $ ($ h_z $) the external magnetic field along the x(z)-direction and $ E $ denotes the strength of the D-M interaction which we interpret as an external electric field $E$ along the $y$ direction coupling to the polarization operator: $P_i^y \equiv S_{i}^{y} S_{i+1}^{x}-S_{i}^{x}S_{i+1}^{y}$(We have taken the direction along the chain to be the $x$ direction, $\vec e_{ij}= \vec x$).
The $XXZ$ model with periodic boundary conditions(PBC) and in a pure longitudinal magnetic field is  known to be exactly solvable by the Bethe ansatz method ~\cite{yang1, *yang2, *yang3}. For the anisotropy parameter $\Delta >1$, the longitudinal magnetic field $h_z$ leads to a sequence of three different phases with increasing field strength: (i) a gapped antiferromagnetic phase with N\'{e}el order along the $z$ direction for $0<h_z<h_{c1}$; (ii) critical Luttinger liquid behaviour for $h_{c1}<h_z<h_{c2}$ and (iii) a fully saturated phase for $h_z>h_{c2}$. The critical fields $h_{c1}$ and $h_{c2}$ can be obtained from the Bethe ansatz solution(for periodic boundary conditions) to be 
\begin{equation}
h_{c1} = \sinh \eta \sum_{k=-\infty} ^\infty \frac{(-1)^k}{\cosh k \eta};\quad h_{c2}= 1 + \Delta
\label{eq:ba_hcr}
\end{equation}
where $\cosh \eta = \Delta$. \\
The $XXZ$ model in a transverse field is not integrable and therefore not amenable to an exact Bethe ansatz solution. For the XXZ model in a pure transverse field, the phase diagram has been obtained by studying the model through diagonalization and DMRG methods~\cite{mori, hieida, dkol} as well as approximate analytic methods~\cite{dkol}. Mean field studies and exact diagonalization studies on small length chains~\cite{dkol} showed that two different phases are possible: the  antiferromagnetic phase with N\'{e}el order along the $z$-direction for $h_x < h_{c3}(\Delta)$ and the magnetically saturated phase (along $x$-direction) for $h_x >h_{c3}(\Delta)$ where $h_{c3}(\Delta)$ is the critical field. ($h_{c3}(\Delta) \approx \sqrt{2(1+\Delta)}$, the classical value)~\cite{dkol}. The effect of both longitudinal and transverse magnetic fields were studied using the mean field approximation~\cite{dmitri-krivnov} for a chain with periodic boundary conditions (PBC) but as far as we are aware, there are no reports of exact phase diagram studies of the anisotropic model ($\Delta >1$) in the presence of both longitudinal and transverse fields. 
In the presence of the electric field and in a pure longitudinal field (i.e. $h_x=0$), the electric field term can be 'gauged' away by a rotation transformation of the spin operators: $S_m^{\pm} \rightarrow \exp{(\pm i m \alpha)}, \quad \tan \alpha = E/J$. This leads to a transformed XXZ Hamiltonian with parameters $\tilde J = \sqrt{ J^2 + \Delta ^2}$ and $\tilde \Delta = \frac{\Delta}{\sqrt{1 + E^2}}$ and with twisted boundary conditions. However such a rotation transformation is not possible when $h_x \neq 0$ and hence the problem cannot be solved using Bethe Ansatz.\\

\begin{figure}[!htpb]
\includegraphics[width=3.0in]{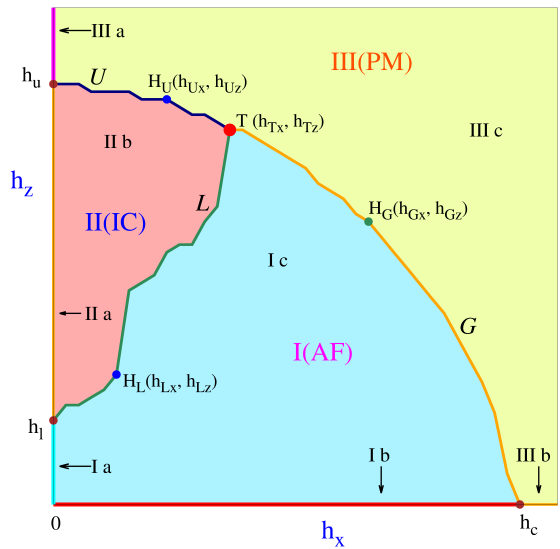}
\caption {(Color online) The schematic of the phase diagram obtained by DMRG for a chain of length $L=64$ and with open boundary conditions. Here $\Delta =4.5$ and $E=0$. Region $I$ depicts a gapped antiferromagnetic (AF) phase, region $II$ a gapless incommensurate (IC) phase and the region $III$, a gapped paramagnetic (PM) phase. These have been described in the text.}
\label{fig:phase_fields_0E}
\end{figure}
\begin{figure}[!htpb]
\includegraphics[width=3.0in]{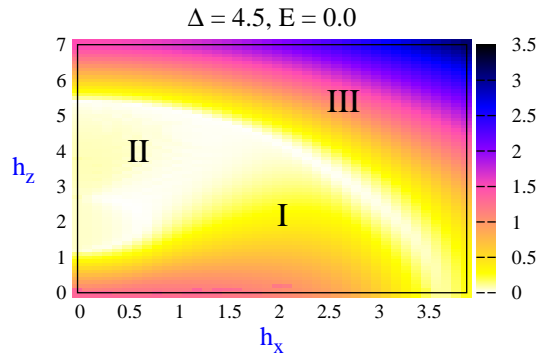}
\caption {(Color online) The $h_x-h_z$ dependence of the energy gap in units of $J(=1)$ (white depicts zero gap). The gapped ($I$ and $III$) and the gapless ($II$) regions are clearly seen. For calculating the energy eigenvalues and the gaps, we set the maximum number of states to $70$ and the number of sweeps to $10$, from which we obtained a truncation error of the order of $10^{-15}$.}
\label{fig:gap_0E}
\end{figure}

In this work, we study the model(Eq.~\ref{eq:ham_dm}) using the numerical DMRG method. We have obtained the energy gap and also studied the behaviour of various physical quantities like the uniform and staggered magnetization, spin correlations, electric polarization, etc.  The numerical calculations have been carried out using the ALPS DMRG application~\cite{ALPS-2}. Further, we restrict to the case $\Delta >1$ and have used open boundary conditions(OBC).  We start by discussing the model (Eq.~\ref{eq:ham_dm}) in the absence of the electric field. From the behaviour of various physical quantities, we identify three different phases: a gapped N\'{e}el antiferromagnetic (AF) phase, a gapped paramagnetic (PM) phase and a gapless critical incommensurate Luttinger (IC) phase  depending on the relative strengths of the longitudinal and transverse magnetic fields as shown in the schematic phase diagram in Fig.~\ref{fig:phase_fields_0E}.
The N\'{e}el and Luttinger phases are separated by the transition curve $L$ while the Luttinger and paramagnetic phases are separated by the transition curve $U$. The transition curve $G$ separates the N\'{e}el and the paramagnetic phase. In Fig. ~\ref{fig:phase_fields_0E}, critical points $H_L$ lying on the transition curve $L$ denote  the critical transverse and longitudinal fields $(h_{Lx}, h_{Lz})$ while points $H_U$ on the transition curve $U$ represents the critical transverse and longitudinal fields $(h_{Ux}, h_{Uz})$. A critical point $h_G$ on the transition curve $G$ denotes the critical transverse and longitudinal fields $(h_{Gx}, h_{Gz})$.  At field strengths $(h_{Tx}, h_{Tz})$, there is a tricritical point $T$ between the three phases. For transverse field strengths, $0 \leq h_x< h_{Tx}$, the longitudinal field $h_z$ gives rise to a sequence of three different phases: a gapped AF phase with N\'{e}el order along both $z$ and $x$ direction for $ 0\leq h_z< h_{Lz}$, a gapless Luttinger phase for $h_{Lz} < h_z< h_{Uz}$ and a magnetically saturated phase for $h_z >h_{Uz}$. The dependence of the lower critical longitudinal field, $h_{Lz}$, and the upper critical longitudinal field, $h_{Uz}$ on the transverse field can be determined from the transition curves $L$ and $U$ respectively. At zero transverse field, $h_{Lz}=h_l$ and $h_{Uz} =h_{u}$.  The lower critical longitudinal field increases with the transverse field, i.e., $h_{Lz}(h_x \neq 0) > h_l$ while the upper critical longitudinal field $h_{Uz}$  decreases with the transverse field or $h_{Uz}(h_x\neq 0) < h_u$. For transverse fields, $h_x>h_{Tx}$, two phases are possible: a gapped AF phase with N\'{e}el order for $0 \leq h_z <h_{Gz}$, and a magnetically saturated phase for $h_z >h_{Gz}$. The longitudinal field dependence of the critical field, $h_{Gx}$ separating the N\'{e}el and magnetically saturated phases is determined by the transition curve $G$. Also, $h_{Gx}(h_z=0)= h_c= h_{c3}(\Delta)$.

\begin{figure*}[!htpb]
\includegraphics[width=3.0in]{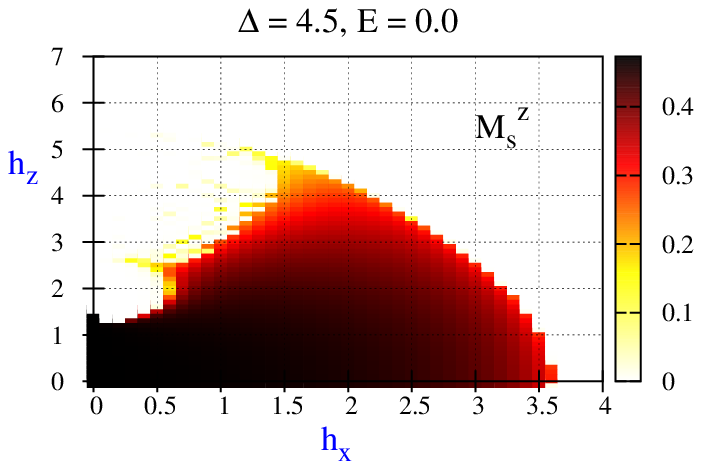}{(a)}
\includegraphics[width=3.0in]{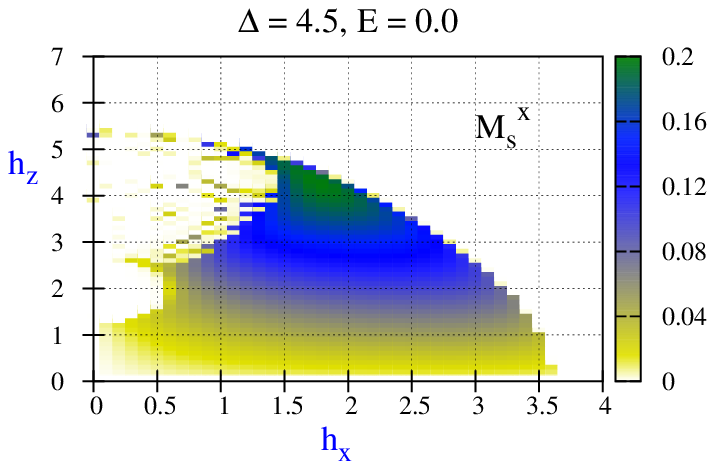}{(b)}
\includegraphics[width=3.0in]{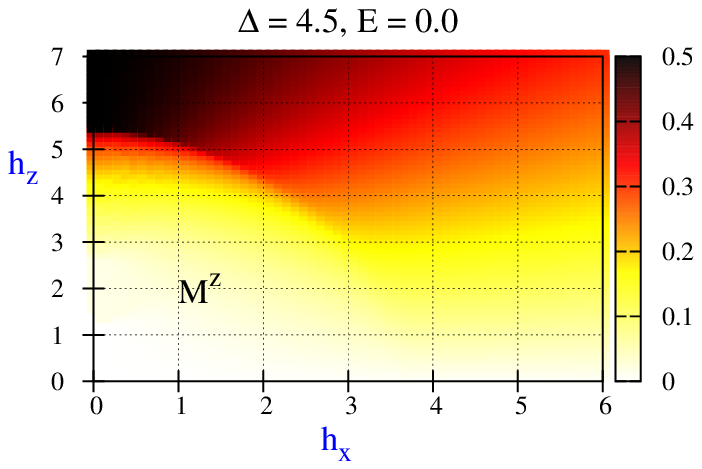}{(c)}
\includegraphics[width=3.0in]{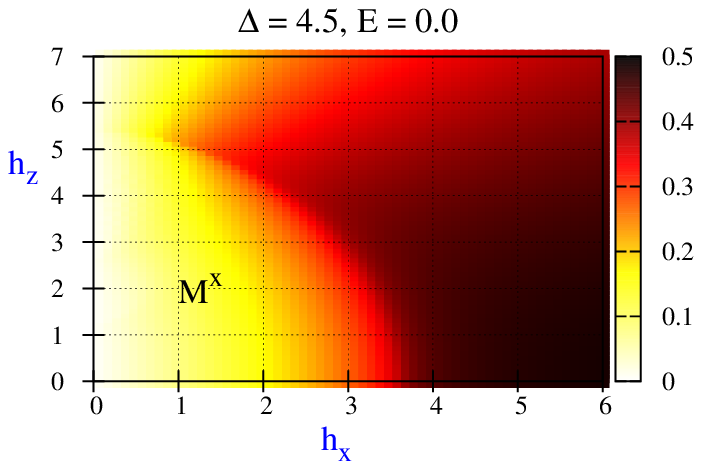}{(d)}
\caption {(Color online) The $h_x-h_z$ dependence of (a) staggered magnetization $M_s^z$ (b) $M_s^x$ ($M_s^x = 0$ along $h_z = 0$), the uniform  magnetization (c) $M^z$ and (d) $M^x$, obtained by DMRG for a chain of length $L=64$ and with open boundary conditions. Here again, $\Delta =4.5$ and $E=0$. For calculating the observables like the magnetization, polarization, etc, we used up to a maximum of $70$ states, the number of sweeps was set to $10$, and we kept a truncation tolerance of $10^{-6}$.}
\label{fig:M_Ms_0E}
\end{figure*}

The above three phases have been identified from the numerical results for the energy gap and the various magnetic orders. We show the $h_x-h_z$ dependence of the energy gap in Fig.~\ref{fig:gap_0E}. Here, the first excitation determines the gap for non-degenerate ground state and the second excitation determines the gap for the degenerate ground state. The $h_x-h_z$ dependence plots for the staggered magnetization $M_s^z$ and $M_s^x$ along the $z$ and $x$ direction are shown in Figs. ~\ref{fig:M_Ms_0E}(a) and ~\ref{fig:M_Ms_0E}(b), respectively while the corresponding plots for the uniform magnetization $M^z$ and $M^x$ are shown in Fig.~\ref{fig:M_Ms_0E}(c) and Fig.~\ref{fig:M_Ms_0E}(d) respectively. 
The uniform magnetization, $M^a$ and the staggered magnetization, $M^a_s$ have been defined as:
\begin{equation}
M^a = \dfrac{1}{N}\sum_{i=1}^{N}\langle S_{i}^{a}\rangle;\quad a = x,y,z
\end{equation}
\begin{equation}
M^a_{s} = \dfrac{1}{N}\sum_{i=1}^{N} (-1)^{i+1} \langle S_{i}^{a}\rangle;\quad a = x,y,z
\end{equation}

\begin{figure*}[!htpb]
\includegraphics[width=3in]{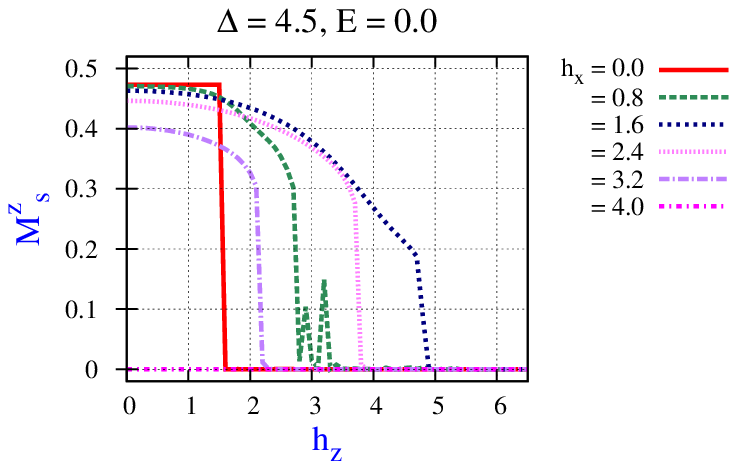}{(a)}
\includegraphics[width=3in]{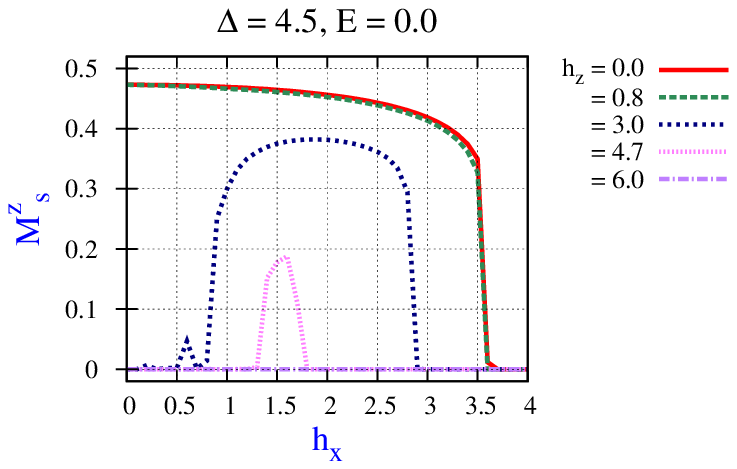}{(b)}
\includegraphics[width=3in]{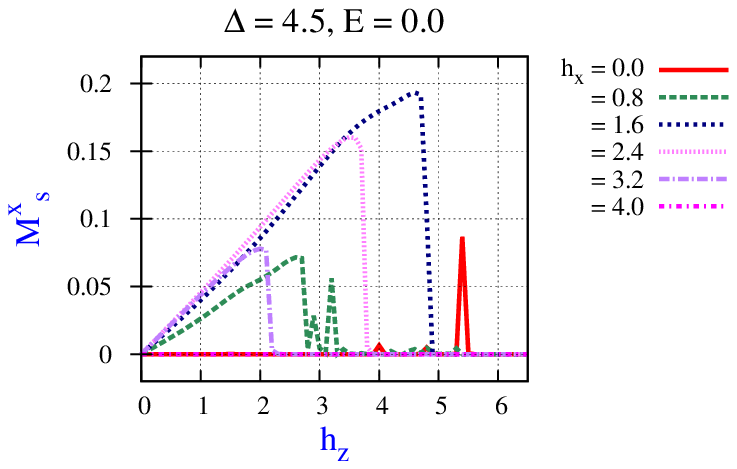}{(c)}
\includegraphics[width=3in]{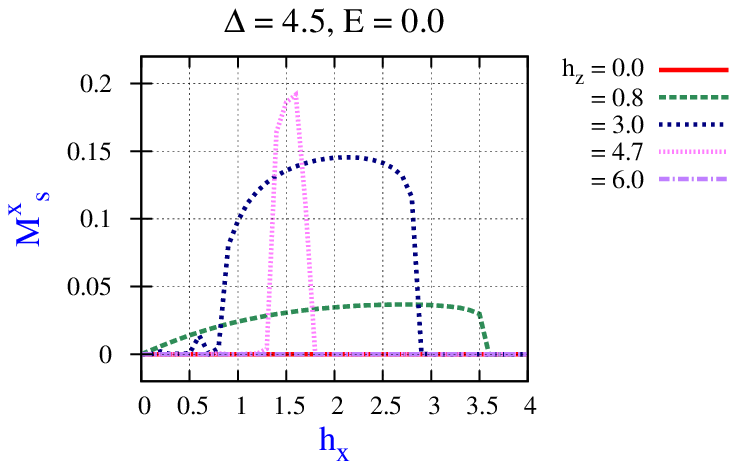}{(d)}
\caption{(Color online) The dependence of the staggered magnetization on $h_z$ and $h_x$ in the absence of the electric field. (a) $M_s^z$ as a function of $h_z$, for different values of $h_x$. (b) $M_s^z$ as a function of $h_x$, for different $h_z$. (c) $M_s^x$ as a function of $h_z$, for different $h_x$. (d) $M_s^x$ as a function of $h_x$, for different $h_z$. All other parameters are as in Fig. \ref{fig:M_Ms_0E}.}
\label{fig:stag_mag_0E}
\end{figure*}

\begin{figure*}[!htpb]
\includegraphics[width=3in]{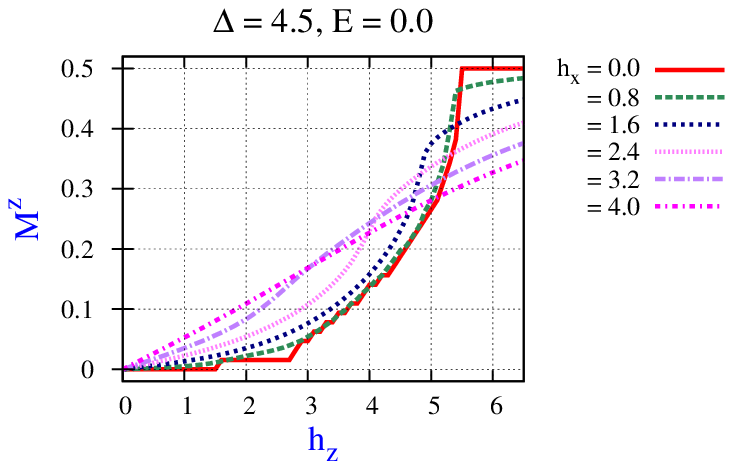}{(a)}
\includegraphics[width=3in]{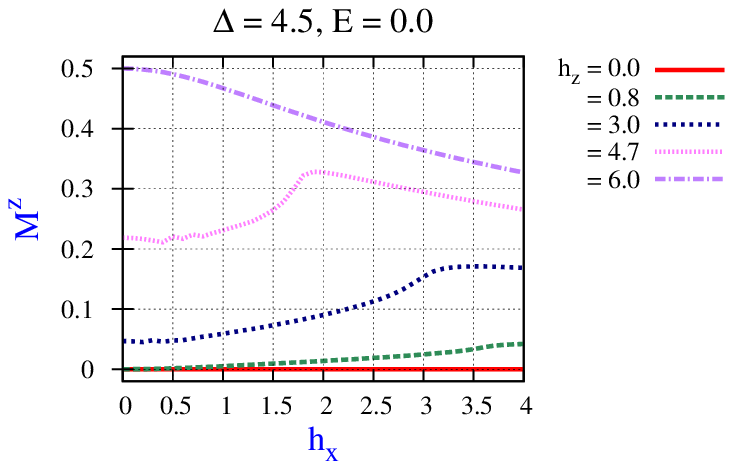}{(b)}
\includegraphics[width=3in]{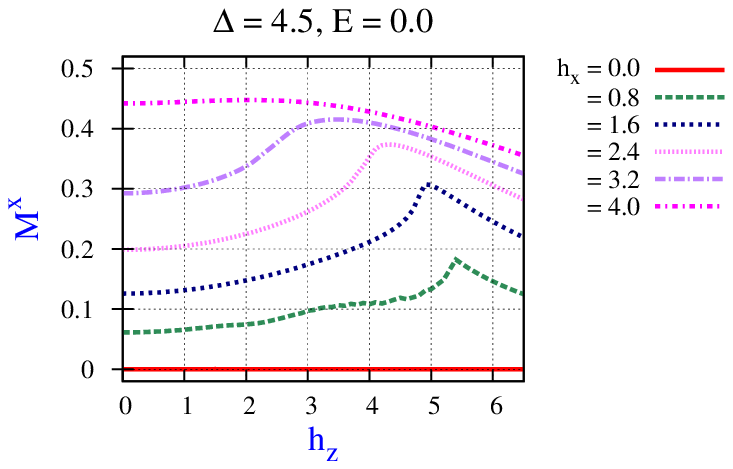}{(c)}
\includegraphics[width=3in]{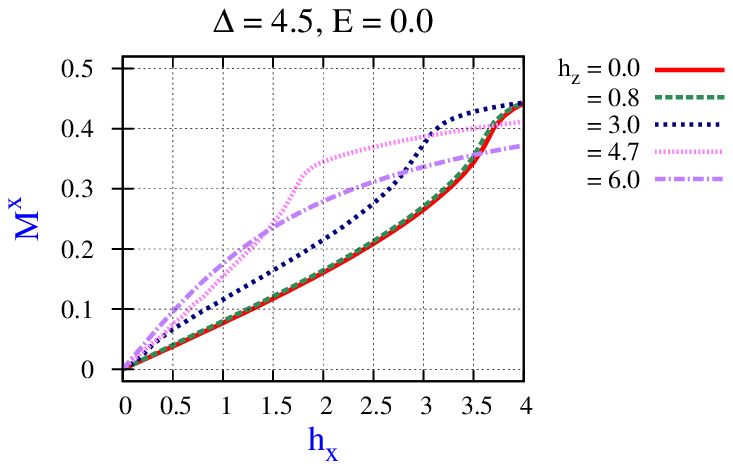}{(d)}
\caption{(Color online) The dependence of the uniform magnetization on $h_z$ and $h_x$ in the absence of the electric field. (a) $M^z$ as a function of $h_z$, for different values of $h_x$. (b) $M^z$ as a function of $h_x$, for different $h_z$. (c) $M^x$ as a function of $h_z$, for different $h_x$. (d) $M^x$ as a function of $h_x$, for different $h_z$. All other parameters are as in Fig. \ref{fig:M_Ms_0E}.}
\label{fig:uni_mag_0E}
\end{figure*}

We distinguish between the three different regions shown in Fig.~\ref{fig:phase_fields_0E} as follows:\\

\noindent (i) Region $I$ corresponds to a gapped phase with antiferromagnetic(AF) N\'{e}el order (non-zero $M_s^z$) along the $z$ direction as can be seen from the plot for the gap shown in Fig.~\ref{fig:gap_0E} and the plot for the staggered magnetization $M_s^z$  along the $z$ direction shown in Fig.~\ref{fig:M_Ms_0E}(a). We further distinguish between three subregions in this phase as described below:

 Subregion $Ia$ corresponds to the case of a pure longitudinal magnetic field. From the plots for the energy gap and staggered magnetization $M_s^z$ (Fig.~\ref{fig:gap_0E} and Fig.~\ref{fig:M_Ms_0E}(a)), it can be seen that here, the lower critical field $h_{Lz} =h_l \sim \frac{1}{2} h_{c1}$ where $h_{c1}$ is the lower critical field as given by the Bethe ansatz relation for PBC (Eq.~\ref{eq:ba_hcr}). This is because we are considering the chain with OBC. This subregion corresponds to an AF phase with order only along the $z$ direction as can be seen from Figs.~\ref{fig:M_Ms_0E}(a) and (b). Also, it can be seen from Figs.~\ref{fig:M_Ms_0E}(c) and (d),that both $M^z$ and $M^x$ are zero in this subregion. 
 
 Subregion $Ib$ corresponds to the case of a pure transverse magnetic field, representing an AF phase with order only along the $z$ direction as can be seen from Figs. ~\ref{fig:M_Ms_0E}(a) and (b). Also, from Figs.~\ref{fig:M_Ms_0E}(c) and (d), it can be seen that $M^x$ is finite in this subregion which tends towards saturation at $h_x=h_c$. However, $M^z$ is zero in this subregion. 

 The subregion $Ic$ corresponds to the case with both longitudinal and transverse fields, representing an AF phase with staggered order along both $z$ and $x$ direction as observed from the plots for the staggered magnetization(Fig.~\ref{fig:M_Ms_0E}(a) and (b)). The staggered order along the $x$ direction arises due to the oscillations in the local magnetization $\langle S_{i}^{x}\rangle$ about a nonzero mean value. It can also be seen from Fig.~\ref{fig:M_Ms_0E}(a) and (b) that the staggered orders in both the $z$ and $x$ directions are nearly equal in the region near the tricritical point $T$. Further in this case, where both longitudinal and transverse fields are present, there is a finite uniform magnetization along both $z$ and $x$ directions as can be seen from Figs.~\ref{fig:M_Ms_0E}(c) and ~\ref{fig:M_Ms_0E}(d). 

\noindent (ii) The region $II$ corresponds to a gapless incommensurate (IC) Luttinger liquid phase. The energy gap vanishes in this region as can be seen from Fig.~\ref{fig:gap_0E}. From Fig.~\ref{fig:M_Ms_0E}(a) and (b), it can be seen that the staggered magnetization along $z$ and $x$ direction are both  zero in this region. We do not show this but the local magnetization $\langle S_i^z \rangle$ and $\langle S_i^x \rangle$ show (in)commensurate oscillatory behaviour. \\
We distinguish two cases here:

Subregion $ IIa $ corresponds to the case of a pure longitudinal field. We can see from  Fig. ~\ref{fig:M_Ms_0E}(c) and (d) that there is a finite uniform magnetization $M^z$ for $h_z > h_l$ which saturates at the upper critical field $h_{Uz}= h_{u}=1 +\Delta$. The uniform magnetization $M^x$ is zero in this subregion.  

Subregion $ IIb $ corresponds to the case where both longitudinal and transverse fields are present($h_{Lz} <h_z<h_{Uz}$, $ 0 < h_x < h_{Tx} $). In this subregion, both $M^z$ and $M_x$ are finite as can be seen from Fig.~\ref{fig:M_Ms_0E}(c) and (d).

\noindent (iii) Region $III$ is a gapped phase described by a magnetically saturated state with uniform magnetization along the direction of the applied field. The energy gap can be seen from Fig. ~\ref{fig:gap_0E}. Here we distinguish between three subregions:

  Subregion $IIIa$ corresponds to the case of a pure longitudinal field ($ h_x = 0 $, $ h_z > h_{u} $). In this subregion, the uniform magnetization $M^z$ is fully saturated as can be seen from Fig.~\ref{fig:M_Ms_0E}(c). 
  
Subregion $IIIb$ corresponds to the case of a pure transverse field. In this subregion, the uniform magnetization is along the $x$ direction as can be seen from Fig.~\ref{fig:M_Ms_0E}(d). However, $M^x$ does not saturate completely because of the competition between the N\'{e}el ordering due to the Ising anisotropy along the $z$-direction on one hand and the quantum fluctuations caused by the effect of the strong transverse field in the easy-plane, on the other. 

Subregion $ IIIc $ corresponds to the case where both longitudinal and transverse magnetic fields are present. As can be seen from Figs. ~\ref{fig:M_Ms_0E}((d) and (e)), there is uniform magnetization along both the $ z $- and the $ x $-directions, i.e, along the direction of the applied field, which tends towards saturation at very large fields. 

\begin{figure*}[!htpb]
\includegraphics[width=3in]{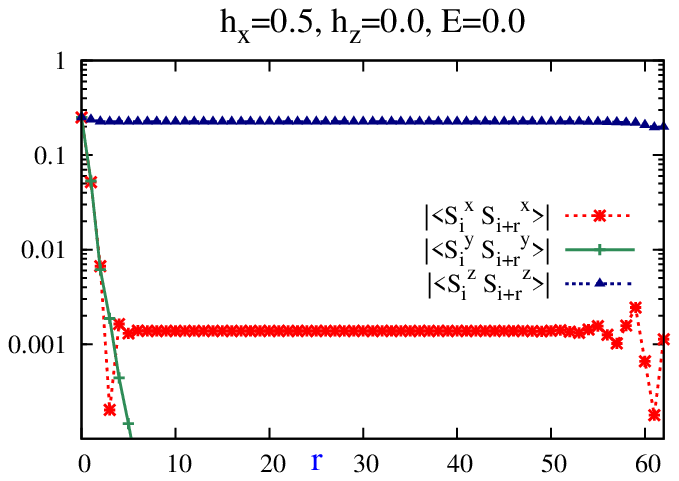}{(a)}
\includegraphics[width=3in]{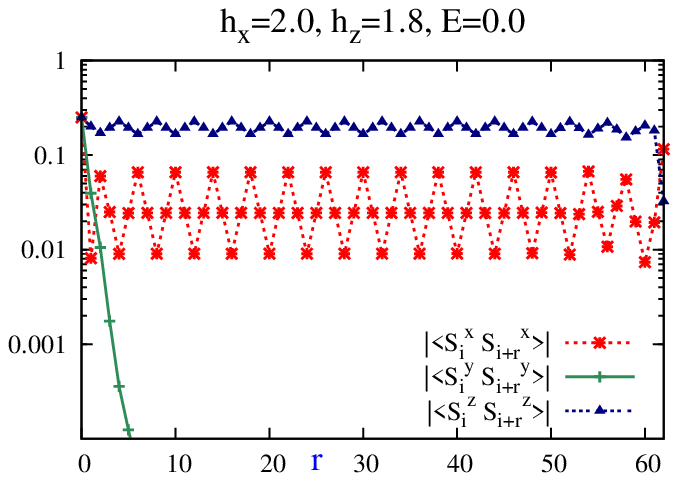}{(b)}
\includegraphics[width=3in]{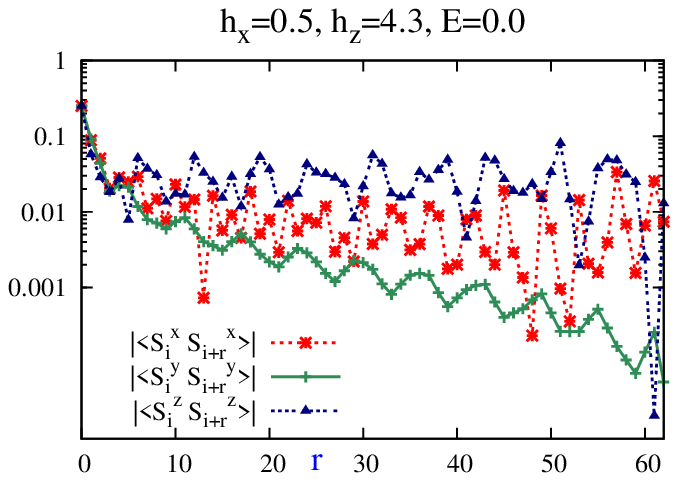}{(c)}
\includegraphics[width=3in]{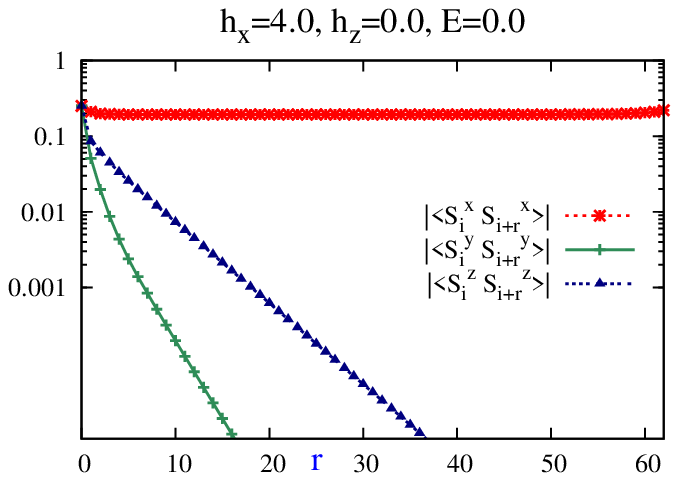}{(d)}
\caption{(Color online) Semilog plots for the spin correlation decays as a function of the distance for $L = 64$, $\Delta = 4.5$ and $E = 0.0$, for representative values of $h_x-h_z$ in the different phases. We only plot here the absolute values. Panel (a) shows the long range (AF) order of the longitudinal spin correlations in subregion $Ib$ ($h_x=0.5$, $h_z=0.0$) of the N\'{e}el phase. A small induced paramagnetic order in the tranverse $x$ component spin correlations can also be observed. Panel (b) describes the effect of both transverse and longitudinal field on the spin correlations in subregion $Ic$ ($h_x=2.0$,$h_z=1.8$) of the N\'{e}el phase. There are additional fluctuations in the paramagnetic $x$ spin correlations due to the longitudinal field. Panel (c) shows the absence of long range order in the subregion $IIb$ ($h_x=0.5$, $h_z=4.3$) of the critical incommensurate phase while panel (d) shows the induced long range ferromagnetic order for the spin correlations in the subregion $IIIb$ ($h_x=4.0$, $h_z=0.0$) of the magnetically saturated phase.}
\label{fig:CDL_v2}
\end{figure*}
The behaviour of the uniform and staggered magnetization in the different phases and the nature of the transitions between the phases can be understood better by studying their dependence on the longitudinal (transverse) field for fixed values of transverse (longitudinal) field. In Figs.~\ref{fig:stag_mag_0E}(a), we plot $M_s^z$  as a function of $h_z$ for fixed $h_x$.  $M_s^z$ is non-zero only in the N\'{e}el phase (region $I$). For a pure longitudinal field, we see that $M_s^z$ is a constant which drops sharply to zero at the lower longitudinal critical field $h_{l}$. In the presence of the transverse field, the magnitude of $M_s^z$ decreases, also the drop to zero at the lower longitudinal critical field $h_{Lz}$ gets smoothened. We show the $h_x$ dependence of $M_s^z$ in Fig.~\ref{fig:stag_mag_0E}(b) from which we find a non-zero value only for longitudinal fields $h_z<h_{Tz}$. From Fig.~\ref{fig:stag_mag_0E}(a) and (b), we also observe that for small transverse and longitudinal fields $M_s^z - \text{const} \sim - a h_x^2 -  b h_z^2$.
In Figs.~\ref{fig:stag_mag_0E}(c) and (d), we plot $M_s^x$ as a function of $h_x(h_z)$ for fixed $h_z(h_x)$. It can be seen from the figure that $M_s^x$ is non-zero for longitudinal fields $h_z<h_{Tz}$ and shows a linear dependence on the longitudinal field, i.e, $M_s^x \sim h_z$. For very small transverse and longitudinal fields, it can be also seen from Figs.~\ref{fig:stag_mag_0E}(c) and (d) that $M_s^x \sim h_z h_x$. 
Further, we see from Fig.~\ref{fig:stag_mag_0E} (a) and (c) that $h_{Lz}$ increases with the transverse field for $h_x<h_{Tx}$. For transverse fields $h_x>h_{Tx}$, the longitudinal critical field $h_{Gz}$ can be seen to decrease with increasing transverse field. Figs. ~\ref{fig:stag_mag_0E}(b) and (d) show that the lower critical transverse field $h_{Lx}$ increases with the longitudinal field while the upper critical transverse field $h_{Gx}$ decreases with the longitudinal field.\\

We next study the behaviour of the uniform magnetization.
In Fig. ~\ref{fig:uni_mag_0E}(a), we plot the uniform magnetization $M^z$ as a function of the longitudinal field for different (fixed) transverse field values. The small step seen in Fig.~\ref{fig:uni_mag_0E}(a) for $M^z$ at $h_l=h_{c1}/2$  for the case of a pure longitudinal field (subregion $Ia$) is due to the finite size effect arising in chains with OBC. Otherwise, the field dependence of $M^z$ is similar to that for a chain with PBC and in a pure longitudinal field~\cite{yang1, yang2, yang3}: the magnetization vanishes in the N\'{e}el phase, increases monotonically in the Luttinger phase reaching  saturation at the upper critical field $h_{c2}=1 +\Delta$. It shows square root singular behaviour\cite{cabra} in the vicinity of the transitions from the N\'{e}el phase to the Luttinger phase and from the Luttinger phase to the magnetically saturated phase.  In finite transverse fields, $0<h_x <h_{Tx}$ ($h_{Tx}\sim 2.0$ for our case), the behaviour is similar, however $M^z$ is now non-zero  in the N\'{e}el phase. Also, the singular behaviour near the transition from the N\'{e}el phase to the Luttinger phase gets smoothened by the transverse field. For transverse fields, $h_x>h_{Tx}$, $M_z$ increases with $h_z$ monotonically, however it no longer reaches saturation value as can be seen from Fig.~\ref{fig:uni_mag_0E}(a). Further, we observe that in general, for small longitudinal fields, $M^z \sim h_z$. We show the corresponding dependence of $M^z$ on the transverse field in Fig.~\ref{fig:uni_mag_0E}(b). For small transverse fields and for longitudinal fields $h_z<h_{Lz}$, it can be seen that $M^z \sim h_x^2$. For $h_{Lz} <h_z <h_{Uz}$, $M^z$ increases slowly with the transverse field, showing a maximum near the transition from the Luttinger IC phase to the magnetically saturated phase, and then decreases with further increase in $h_x$ as can be seen from Fig.~\ref{fig:uni_mag_0E}(b). For $h_z>h_{Uz}$, $M^z$ decreases with increasing $h_x$. 
From Fig.~\ref{fig:uni_mag_0E}(a) and (b), we conclude that for small transverse and longitudinal fields, $M^z \sim h_z h_x^2$ which is in agreement with mean field theory results~\cite{dmitri-krivnov}. 
In Fig. ~\ref{fig:uni_mag_0E}(c), we plot the uniform magnetization $M^x$ as a function of the longitudinal field for different (fixed) transverse field values, while in Fig.~\ref{fig:uni_mag_0E}(d), we plot the transverse field dependence of $M^x$. From the two plots, we can see that $M^x \sim h_x$ for small transverse fields and that it tends towards saturation in subregions $IIIb$ and $IIIc$. 

We also study the behaviour of the spin-correlation decays in order to confirm the above identification of the different phases. We define the two-spin correlation functions $C^a(n)$ as: 
\begin{equation}
C^a(n) \equiv \langle C^a_{i,i+n}\rangle = \langle S_{i}^{a}S_{i+n}^{a} \rangle; \quad a=x,y,z
\end{equation}
where $n$ is the distance between the two spins. In general, in the gapped regions $ I $ and $ III $, the spin correlations show long range AF order and long range FM order respectively while in the gapless phase $ II $, the spin correlations show an algebraic decay. For a pure longitudinal field, the transverse spin correlations are isotropic in nature because of the spin rotation symmetry about the $z$ axis. In the subregion $Ia$(N\'{e}el phase), the longitudinal spin correlations (along the $z$ direction) show AF staggered long range order while the transverse spin correlations($ x $ and the $ y $ components) decay exponentially to zero. In subregion $IIa$ (IC phase), the spin correlations show a power-law decay. In subregion $IIIa$(magnetically saturated phase), the longitudinal spin correlations show long range FM order while the transverse spin correlations decay exponentially. These results are described in Fig.~\ref{fig:CDL_v2}, where we show the semilog plots of the absolute values of the two spin correlation decays $ |C^a(n)|, a=x,y,z $ as a function of the distance $n$ for representative field values in the various phases focussing on the effect of transverse field. Panel (a) of the figure shows the spin correlation decays for a pure transverse field (subregion $Ib$). It can be seen that the longitudinal spin correlations show long-range (AF) order. The spin correlations along the $x$ direction decay to a finite uniform value while the spin correlations along the $y$-direction decay exponentially to zero (the spin rotation symmetry in the $x-y$ plane is broken due to the transverse field). In the presence of both longitudinal and transverse fields (subregion $Ic$), as can be seen from Fig.~\ref{fig:CDL_v2}(b), both longitudinal and transverse spin correlations along the $x$ direction show long range order with small oscillations about the mean values. The transverse spin correlations along the $ y $ direction decay exponentially. Panel (c) of Fig.~\ref{fig:CDL_v2} depicts the two spin correlation decay behaviour for representative values of longitudinal and transverse fields in the gapless incommensurate phase (subregion $ IIb $). We see here that the transverse and longitudinal spin correlations decay in accordance with a power-law, albeit with fluctuations about the mean decay. Also, the $x-y$ isotropy is broken. In panel (d) of the figure we show the behaviour of the spin correlations for a pure transverse field in the fully saturated phase (subregion $ IIIb $). There is long range ferromagnetic order of the spin correlations along the direction of the applied field, $ h_x $. The $ y $- and the $ z $-correlations decay exponentially. We do not show this but, in the presence of both transverse and longitudinal fields(subregion $IIIc$), the spin correlations show long range FM order along the direction of the applied field.
 
While the above phase diagram is in general consistent with earlier mean field approximation studies ~\cite{dmitri-krivnov}, there are some differences. The phase boundary of the critical phase which we have determined shows that in a small applied transverse field $h_x$, the sequence of phase transitions with increasing longitudinal magnetic field $h_z$ is similar to that in the absence of the transverse field. This is in contrast to what was suggested in Ref.~\cite{dmitri-krivnov} that there is an intermediate AF phase between the critical phase and the saturated phase due to the induced anisotropy operator $\sum_i [S_i^x S_{i+1}^x - S_i^yS_{i+1}^y]$ becoming relevant beyond a certain field strength $h_z >h_{z1}$. We do observe such an induced anisotropy as discussed in the next paragraph, however, it does not lead to any intermediate AF phase. We argue that this is because, at these field strengths, the Zeeman coupling $h_z \sum_i S_i^z$ becomes much larger and hence there is a transition directly from the IC phase to the saturated phase. For higher transverse field strengths ($h_x>h_{Tx}$), with increasing $h_z$, there is only one phase transition from the AF N\'{e}el phase to the fully saturated phase.

The transverse magnetic field also leads to other kinds of order. One such effect is the nematic order $ \langle S_i^+ S_{i+1}^+\rangle $ which arises due to the $x-y$ anisotropy induced by the transverse field. The real and the imaginary parts of the uniform nematic order are defined as~\cite{dagotto-zhang} 
\begin{eqnarray}
\text{Re}\, N_u &= &\frac{1}{N} \sum_i \langle S_i^x S_{i+1}^x  - S_i^y S_{i+1}^y \rangle \nonumber \\
\text{Im}\, N_u &= &\frac{1}{N} \sum_i \langle S_i^x S_{i+1}^y  + S_i^y S_{i+1}^x \rangle
\end{eqnarray}
\begin{figure}[!htpb]
\includegraphics[width=3.0in]{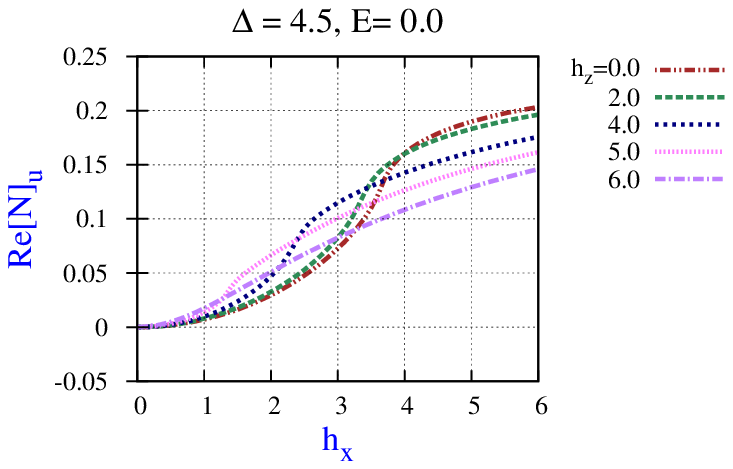}{(a)}
\includegraphics[width=3.0in]{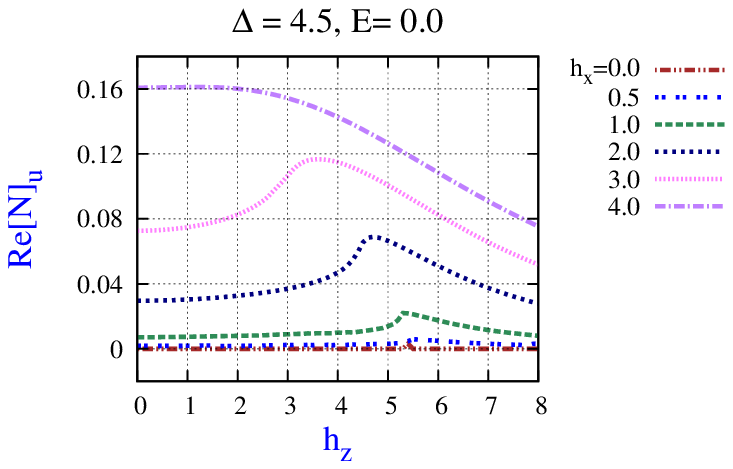}{(b)}
\includegraphics[width=3.0in]{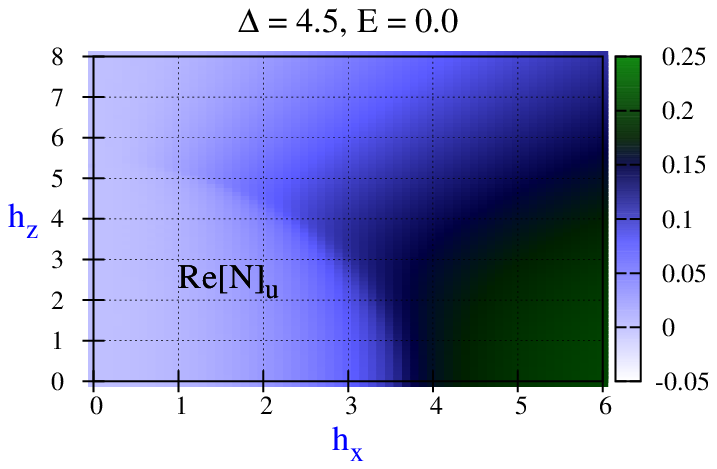}{(c)}
\caption{(Color online) The behaviour of the uniform average of the real part of the nematic correlations $ Re[N]_u $ as a function of the longitudinal and transverse fields and in the absence of the electric field. The different panels describe: (a) transverse field dependence of $ Re[N]_{u}$ for fixed longitudinal field values; (b) longitudinal field dependence of $ Re[N]_{u}$ for fixed transverse field values; (c) the combined $h_x-h_z$ dependence of $ Re[N]_u $ . $Re[N]_u$ is identically zero for $h_x=0$.}
\label{fig:nematic_0p0E}
\end{figure}

In the absence of the electric field, the transverse field induces only the real part of the uniform nematic order $ Re[N]_u $. 
From Fig.~\ref{fig:nematic_0p0E}(a), where we show the dependence of the uniform average of the real part $ Re[N]_u $ on the transverse field for fixed $h_z$, we can see that the nematic order increases monotonically with increasing transverse field tending to saturation in the magnetically saturated (along $x$ direction) phase. We plot the longitudinal dependence of $ Re[N]_u $ in Fig.~\ref{fig:nematic_0p0E}(b) for fixed transverse fields. We observe that for small transverse fields, the nematic order increases slowly with $h_z$, reaches a maximum near the transition to the magnetically saturated (along $z$ direction) phase, and then decreases in the fully saturated phase. For larger transverse fields, the nematic order is large for small $h_z$ and then eventually decreases with increasing $h_z$. The combined effects of both the transverse and longitudinal fields on $Re [N]_u$ are shown in Fig.~\ref{fig:nematic_0p0E}(c). It can be seen from the figure that the nematic order is small in the N\'{e}el and Luttinger (IC) phases, increases with the transverse field and attains its maximum value  in the magnetically saturated phase (subregions $IIIb$ and $IIIc$). 

The transverse field also induces a small real part of the staggered nematic order(with magnitude two orders smaller than the uniform part) where the real and imaginary parts of the staggered nematic order are defined as 
\begin{eqnarray}
\text{Re}\, N_s &=& \frac{1}{N} \sum_i (-1)^i\langle S_i^x S_{i+1}^x  - S_i^y S_{i+1}^y \rangle \nonumber \\
\text{Im}\, N_s &=& \frac{1}{N} \sum_i (-1)^i \langle S_i^x S_{i+1}^y  + S_i^y S_{i+1}^x \rangle
\end{eqnarray}
The presence of both the uniform and staggered nematic orders suggests that the transverse field induces a bond-dimerization~\cite{dagotto-zhang}. Further, the breaking of the spin $U(1)$ rotational symmetry by the transverse field also leads to a small nematic order of the type: $N_{u}^{\pm} = \dfrac{1}{N}\sum_{i} \langle (S_{i}^{z}S_{i+1}^{\pm} + S_{i}^{\pm}S_{i+1}^{z}) \rangle $. \\
\begin{figure}[!htpb] 
\includegraphics[width=3.0in]{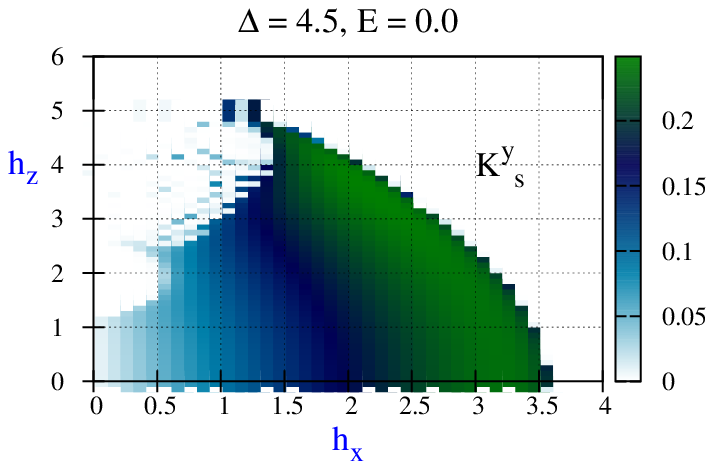}{(a)}
\includegraphics[width=3.0in]{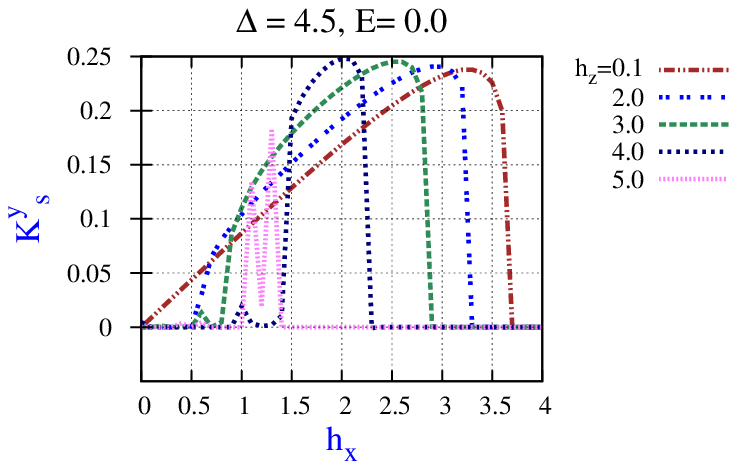}{(b)}
\includegraphics[width=3.0in]{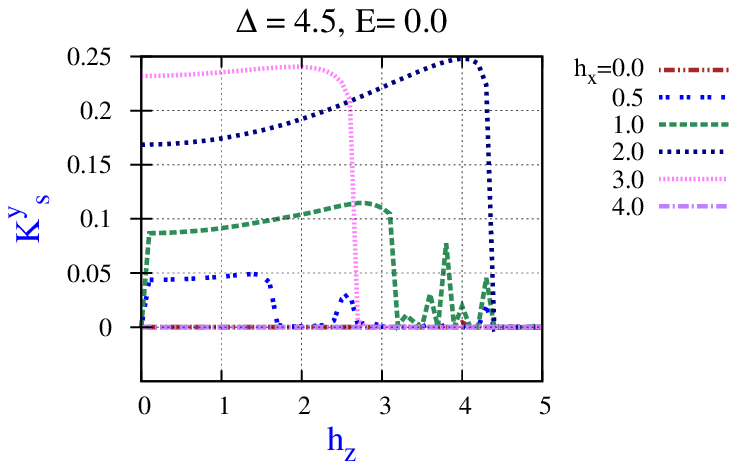}{(c)}
\caption{(Color online) The behaviour of the real part of the staggered vector chirality $ K^y_s$ as a function of the longitudinal and transverse magnetic fields and in absence of the electric field. The different panels describe (a) $h_x-h_z$ dependence of $K^y_s$; (b) transverse field dependence of $ K^y_s$ for fixed longitudinal field values; (c) longitudinal field dependence of $ K^y_s$ for fixed transverse field values. $K^y_s$ is zero for $h_x=0$.}
\label{fig:chirality_0p0E}
\end{figure}

More interestingly, the transverse field also induces a vector chirality in the $x-y$ plane where we define the uniform and staggered in-plane vector chirality as 
\begin{eqnarray}
K_u^{\pm} &=& \dfrac{1}{N}\sum_{i} \langle (S_{i}^{z}S_{i+1}^{\pm}-S_{i}^{\pm}S_{i+1}^{z}) \rangle \\
K_s^{\pm} &=& \dfrac{1}{N}\sum_{i} (-1)^i \langle (S_{i}^{z}S_{i+1}^{\pm}-S_{i}^{\pm}S_{i+1}^{z}) \rangle
\end{eqnarray}
The real and imaginary parts of $K_u^{+}$ can be written as: 
\begin{eqnarray}
K_u^y &\equiv& \text{Re} K_u^{+} = \dfrac{1}{N}\sum_{i} \langle (S_{i}^{z}S_{i+1}^{x}-S_{i}^{x}S_{i+1}^{z}) \rangle  \\
K_u^x &\equiv& -\text{Im} K_u^{+} = - \dfrac{1}{N}\sum_{i} \langle (S_{i}^{z}S_{i+1}^{y}-S_{i}^{y}S_{i+1}^{z}) \rangle
\end{eqnarray}
and similarly for the staggered chirality. The vector chiral order arises due to the competition between the AF ordering due to the Ising anisotropy and the magnetic ordering due to the transverse field $h_x$. From Fig.~\ref{fig:chirality_0p0E}(a), where we show the $h_x-h_z$ dependence of $K^y_s$, it can be seen that there is a large staggered vector chiral order in the N\'{e}el phase. For fixed $h_z$, and in the N\'{e}el phase, the vector chirality $K_{s}^y$ shows a linear dependence on the transverse magnetic field, $ h_x $, as can be seen from Fig.~\ref{fig:chirality_0p0E}(b). Further, we can see from Fig.~\ref{fig:chirality_0p0E}(b), that $K_s^y$ goes to zero sharply at the transition from the N\'{e}el phase to the magnetically saturated phase. The $ h_z $-dependence of $K^y_s$ for fixed $h_x$ is shown in Fig.~\ref{fig:chirality_0p0E}(c). In the N\'{e}el phase, $K_s^y$ shows almost  plateau-like behaviour and then goes to zero in the critical phase.
We also find a small uniform  $K_{u}^y$ (two orders of magnitude smaller than the staggered chiral order)in the N\'{e}el phase. The AF order along the $z$ direction is responsible for the fact that the magnitude of the staggered chiral order $K_s^y$ is much larger than the uniform chiral order $K_u^y$. Within the spin current mechanism, $K^y$ can be considered as an electric polarization along the $z$ direction.

\section{\label{sec:electric} XXZ chain with magnetic and electric fields}

\begin{figure*}[!htpb] 
\includegraphics[width=3.0in]{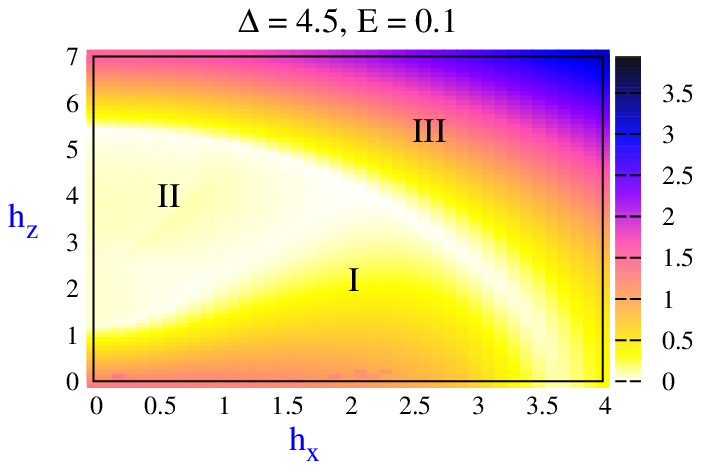}{(a)}
\includegraphics[width=3.0in]{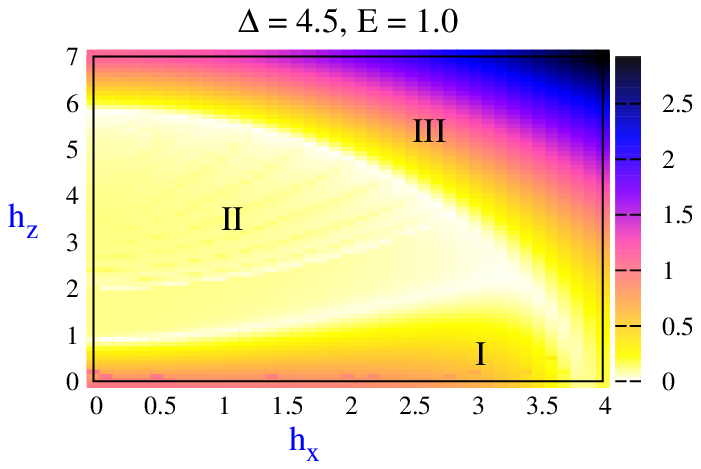}{(b)}
\includegraphics[width=3.0in]{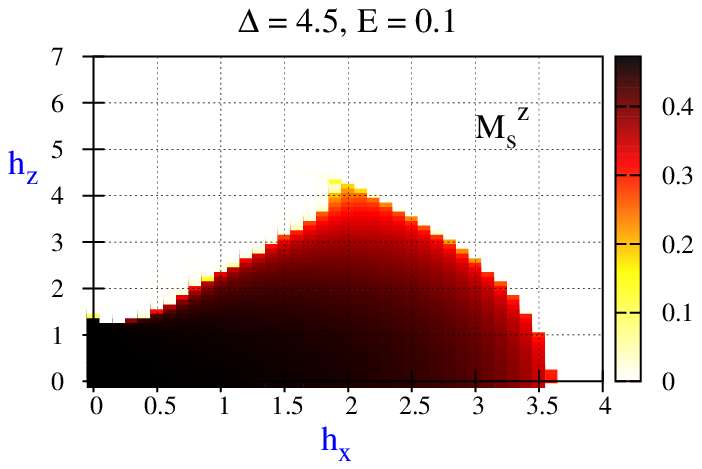}{(c)}
\includegraphics[width=3.0in]{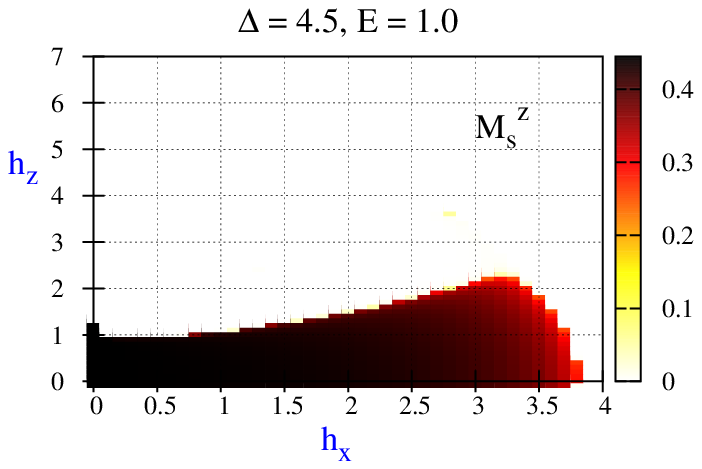}{(d)}
\includegraphics[width=3.0in]{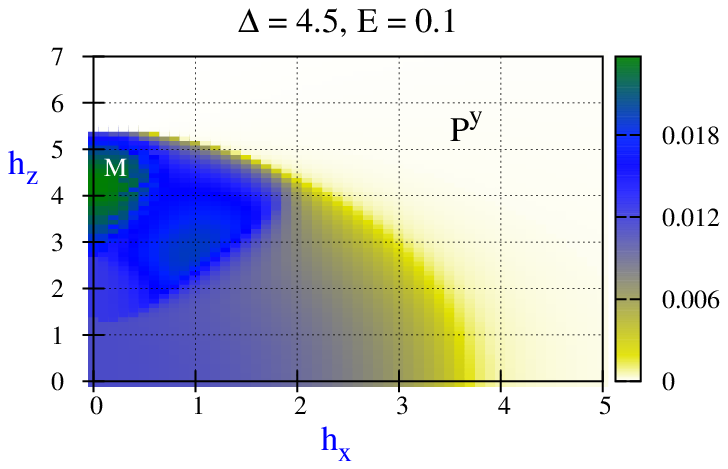}{(e)}
\includegraphics[width=3.0in]{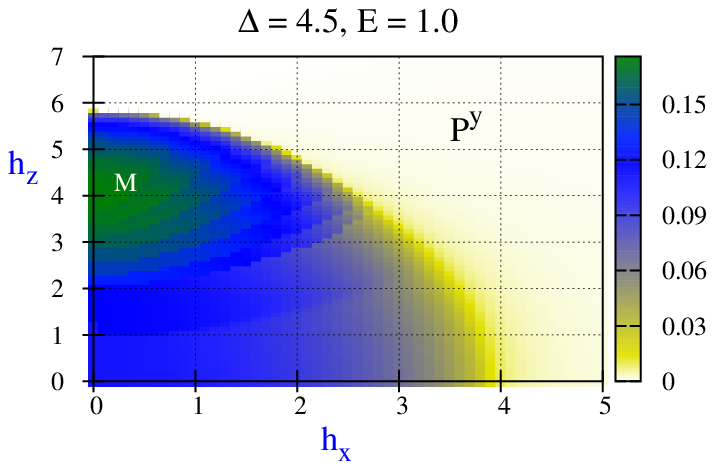}{(f)}
\caption{(Color online) The $h_x-h_z$ dependence of (a),(b) the energy gap; (c),(d) staggered magnetization $M_s^z$; and (e),(f) polarization $P^y$ for $E =0.1$ and $E=1.0$ respectively. All other parameters are as in Fig.\ref{fig:M_Ms_0E}. The white colour depicts zero for all observables.}
\label{fig:gap_diagram_diff_E}
\end{figure*}

In this section, we discuss the combined effect of the external electric field along the $y$ direction and the longitudinal and transverse magnetic fields. We present our numerical DMRG results for the phase diagram and the physical observables when both the longitudinal and transverse magnetic fields are present and there is an electric field as well. The main result we find is that the electric field does not give rise to any new order, it however tends to destroy the AF order and increase the fluctuations of the transverse spin components, leading to a modification of the phase boundaries. This can be seen from Fig.~\ref{fig:gap_diagram_diff_E}(a,b,c,d) where we show the $h_x-h_z$ dependence of the energy gap and the staggered magnetization for two representative values of the electric field $E=0.1$ and $E=1.0$. It can be seen from panels (a) and (b) of the figure that with increasing electric field strengths, there is an increase in the extent of the IC gapless Luttinger liquid phase at the cost of the AF N\'{e}el and the magnetically saturated phases. This can also be corroborated from Figs.~\ref{fig:gap_diagram_diff_E}(c) and (d), where we see that the staggered magnetization $ M_s^z $ is non-zero in a smaller parameter region as compared to that in the absence of the electric field. Also, the magnitude of the order parameter decreases with increasing strength of the electric field.
\begin{figure*}[!htpb]
\includegraphics[width=3.0in]{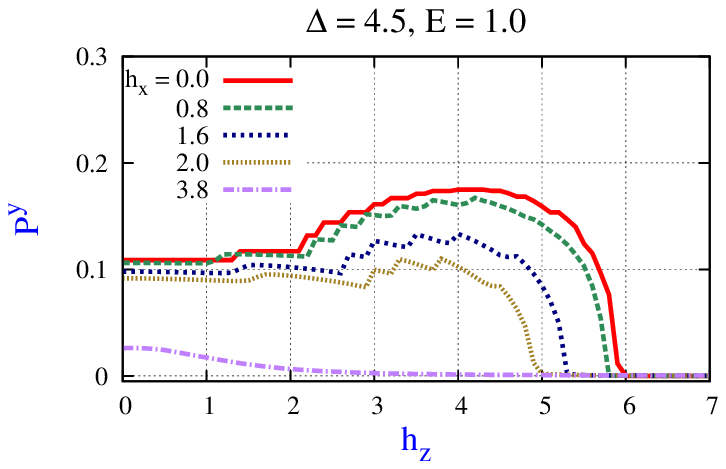}{(a)}
\includegraphics[width=3.0in]{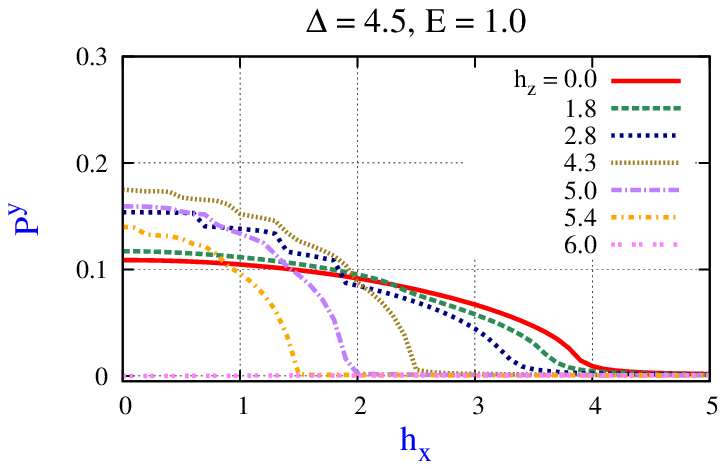}{(b)}
\caption{(Color online) The polarization $P^y$ as (a) a function of the longitudinal field for different transverse field values, and as (b) a function of the transverse field for different longitudinal field values.
The electric field value is set at $E=1.0$. All other parameters are as in Fig.\ref{fig:M_Ms_0E}.}
\label{fig:Py_diff_E}
\end{figure*}
The behaviour of all other quantities like $M^z$, $M^x$, $M_s^x$ is consistent with the above phase diagrams. The spin correlation decay also behave as expected in the different phases. The only difference is that the electric field induces small fluctuations in the spin correlations along the $y$ direction in the critical IC phase.\\

\begin{figure*}[!htpb]
\includegraphics[width=3.0in]{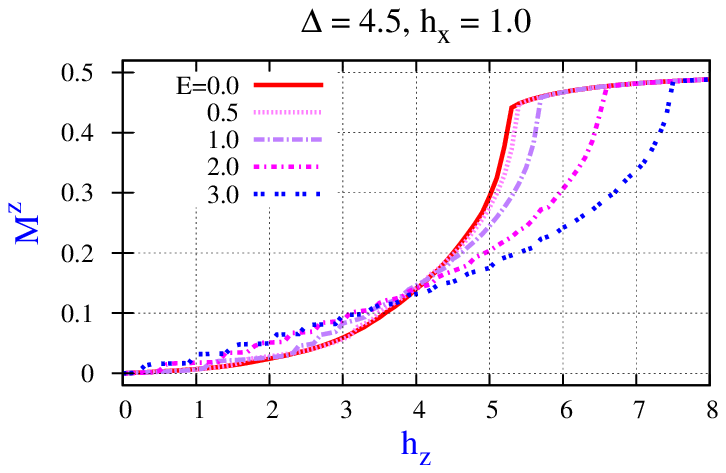}{(a)}
\includegraphics[width=3.0in]{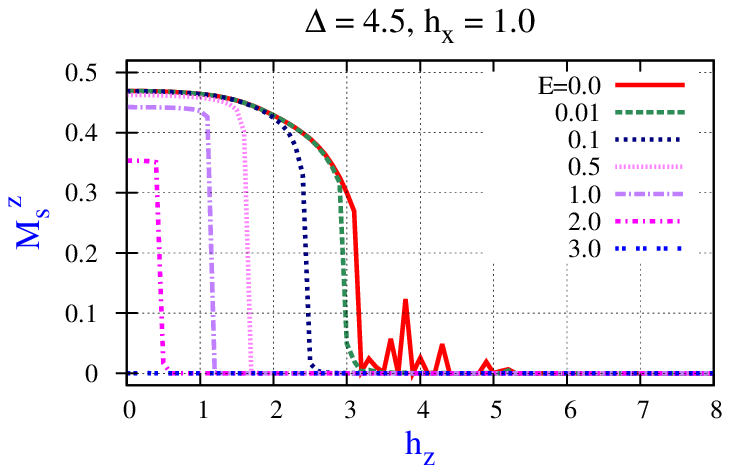}{(b)}
\includegraphics[width=3.0in]{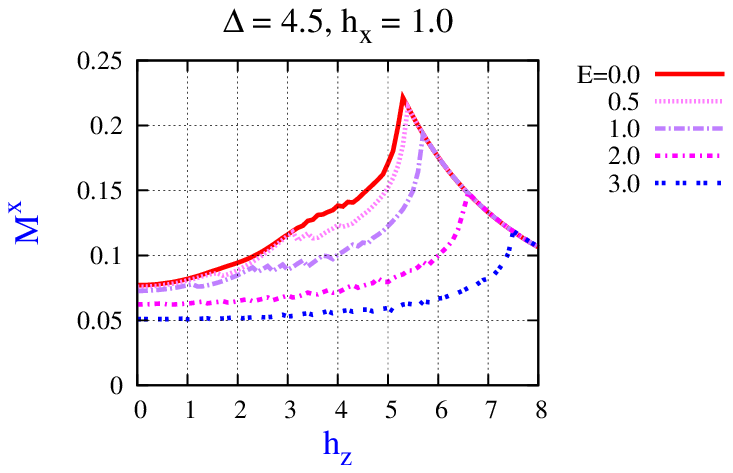}{(c)}
\includegraphics[width=3.0in]{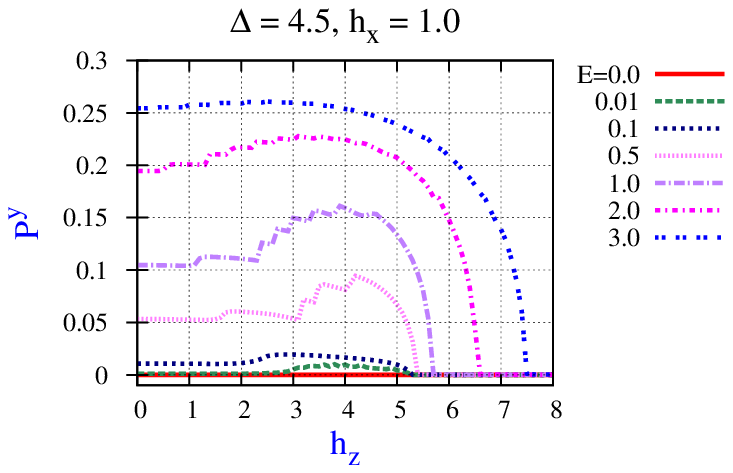}{(d)}
\caption{(Color online) 
The combined effect of the transverse field $ h_x $ and the electric field $E$ on the $h_z$-dependence of the observables (a) $ M^z $, (b) $ M_s^z $, (c) $M^x$ and (d) $ P^y $. Here the transverse field is kept fixed at $ h_x = 1.0 $ and $ E $ takes zero and different non-zero values, all other parameters being as in Fig. \ref{fig:M_Ms_0E}. The effect of $h_x$ (smoothening of the transitions) is suppressed by the electric field, clearly seen in panels (a) and (b).}
\label{fig:mag_pol_hz_E}
\end{figure*}

Besides the uniform and staggered magnetization, in the presence of an electric field, the other observable of interest is the uniform electric polarization $P^y = \dfrac{1}{N}\sum_{i} \langle P_i^y\rangle$. While the polarization $P^y$ is induced by the external electric field ($P^y =0$ when $E=0$), its behaviour is influenced both by the electric and magnetic fields. We show the $h_x-h_z$ dependence of $P^y$ in Fig.~\ref{fig:gap_diagram_diff_E}(e) and (f) for the two representative values of $E=0.1$ and $E=1.0$, respectively. Since the electric polarization is due to the canting of the $x-y$ components of the spins, $P^y$ is significant in the AF N\'{e}el and the IC critical phase i.e., in the regions where the nearest neighbour spin correlations are antiferromagnetic. $P^y$ takes its maximum value in the IC phase (region M of Fig.~\ref{fig:gap_diagram_diff_E}(e, f)). It is very small in the magnetically saturated phase. It can also be seen from the figure that the magnitude of the electric polarization increases with the electric field. 
We can understand the behaviour of the electric polarization in the different phases better by studying the $h_z(h_x)$ dependence of the electric polarization for fixed transverse(longitudinal) field values as shown in Fig.~\ref{fig:Py_diff_E}(a(b)) for $E=1.0$. It can be seen from Fig.~\ref{fig:Py_diff_E}(a), that in a pure longitudinal field($h_x=0$), $P^y$ shows a plateau in the N\'{e}el phase,non-monotonic dependence on $h_z$ in the critical IC phase and then eventually vanishes in the fully saturated  phase. This is consistent with earlier results~\cite{brockmann, pradeep}. The non-monotonic $h_z$ dependence of the polarization in the IC phase can be related to the fact that in the critical spin liquid phase for $\Delta >1$, there is a crossover from an Ising-like phase with dominant incommensurate longitudinal spin correlations to an XY-like phase with dominant staggered transverse spin correlations~\cite{haldane, kimura}. The crossover takes place at a field $h_{l}<h*<h_u$.  We suggest that the non-monotonic behaviour of the polarization in the incommensurate LL  phase is also due to this crossover. The polarization increases with decreasing longitudinal spin correlations and then decreases with increasing transverse spin correlations (The transverse spin correlations also decrease in magnitude but the longitudinal spin correlations go to zero faster and before the transverse correlations do). The maximum value of the polarization is near the field value where the crossover takes place. From Fig.~\ref{fig:Py_diff_E}(b), we see that in the presence of a pure transverse magnetic field, $P^y$ is maximum at zero field and then decreases monotonically with increasing $h_x$ until it vanishes in the magnetically saturated phase at $h_x=h_{Gx}$.
Since a finite transverse field favours alignment of spins along the field direction, this leads to a decrease in the AF spin correlations and hence a decrease in the canting of the spins or the electric polarization. Such a monotonic decrease with increasing $h_x$ is also observed in the presence of finite longitudinal fields in the N\'{e}el phase($0<h_z<h_{Lz}$). This can also be seen from Fig.~\ref{fig:Py_diff_E}(a) where finite transverse leads to a decrease in the magnitude of the polarization. However, the maximum value that $P^y$ can attain, changes non-monotonically for longitudinal fields in the range $h_{Lz}< h_z<h_{Uz}$,  again for reasons described in the previous paragraph.

We can also study the combined effect of the magnetic and electric fields of the various physical observables like the uniform and staggered magnetization and the polarization. In the absence of the transverse field, the longitudinal magnetic field dependence of the staggered magnetization $M_s^z$ and $M^z$ are similar to that shown in the absence of the electric field(Fig.~\ref{fig:stag_mag_0E} and Fig.~\ref{fig:uni_mag_0E})), only the critical field values $h_l$ and $h_u$ change due to the electric field. 
The effect of a transverse field on the $h_z$ dependence of the uniform and staggered magnetization, $M^z, M^x, M_s^z$, and polarization $P^y$ are shown in Fig.~\ref{fig:mag_pol_hz_E}(a, b, c, d) for different $E$ values. It can be seen from Figs.~\ref{fig:mag_pol_hz_E}(a) and (b), that for small electric fields, the uniform ($M^z$) and staggered magnetization($M_s^z$) shows similar qualitative behaviour in the three different phases as in the absence of the transverse field. The main effect of the transverse field is that we now no longer observe the cusp square root singular behaviour near the critical fields $h_{Lz}$ and $h_{Uz}$. Interestingly, it can also be observed from the figures that the  singular behaviour is restored for electric field strength values comparable to that of the transverse field. Finite electric fields also tend to reduce the magnitude of $M^x$ as can be seen from Fig.~\ref{fig:mag_pol_hz_E}(c). The polarization $P^y$ (Fig.~\ref{fig:mag_pol_hz_E}(d)) also shows similar qualitative behaviour as in the absence of the transverse field. Again, for electric fields small compared to the transverse field, there is a smoothening of the cusp singularity near the critical fields. Further, since the transverse field tends to align the spins along the $x$ direction while the electric field tends to rotate the spins in the $x-y$ plane, the magnitude of the electric polarization decreases for electric fields small compared to the transverse fields. For electric field strength values large compared to that of the transverse field, one finds that the magnitude of the polarization does not change very much compared to that in the absence of the transverse field and also the square root singular behaviour near the critical fields is restored. Or in other words, the effect of the transverse field is reduced by large electric fields.

\begin{figure}[!htpb]
\includegraphics[width=3.0in]{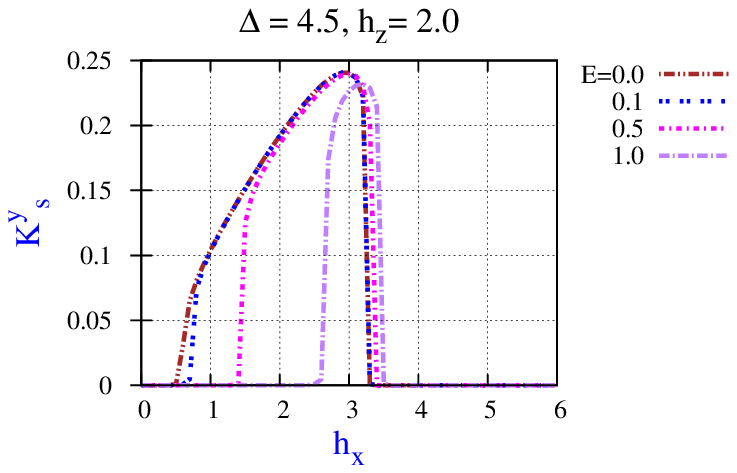}{(a)}
\includegraphics[width=3.0in]{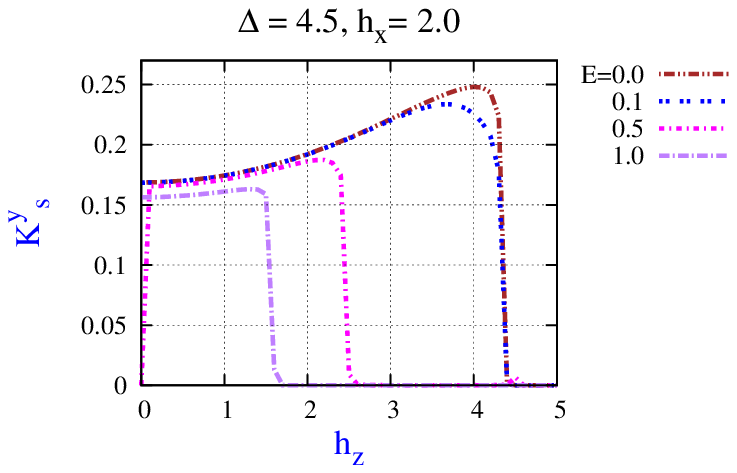}{(b)}
\caption{(Color online) (a) The transverse field dependence of 
$K^y_s$ for fixed longitudinal field value $h_z=2.0$ and for different $E$ values; (b) the longitudinal field dependence of $ K^y_s $ for fixed transverse field value $h_x=2.0$ and for different $E$ values.}
\label{fig:chirality_1p0E}
\end{figure}
We next discuss the effect of the electric field on the nematic order and the vector chirality. Qualitatively, the behaviour of the real part of the nematic order and the vector chirality as a function of the longitudinal and transverse field does not change very much in the presence of the electric field. The main difference is that the $h_x-h_z$ parameter regimes where the nematic and chiral order become significant change due to the modification of the phase boundaries by the electric field. We demonstrate this in Fig.~\ref{fig:chirality_1p0E}. From Fig.~\ref{fig:chirality_1p0E}(a), where we plot the staggered vector chirality as a function of $h_x$ for different $E$ values at a fixed value of $h_z$,  we can see that with increasing electric field strengths,  while larger threshold transverse fields are required to induce the vector chiral order, the functional dependence on the transverse field in the IC phase does not change very much in the presence of the electric field. Electric field strengths large compared to that of the transverse field tend to reduce the chiral order. This can be seen from
Fig ~\ref{fig:chirality_1p0E}(b), where have shown the vector chirality as a function of $h_z$ for different $E$ values at a fixed value of $h_x$. It can be seen from the figure that the magnitude of the chiral order decreases with increasing electric field strengths. Also, larger electric field values tend to restore the sharpness of the transition. 
We do not show this, but in addition, the electric field also induces the imaginary part of both the nematic order and the vector chirality; however, in comparison to their real counterparts, these are two orders of magnitude smaller.
$\text{Im} N_{u(s)}$ is non-zero only in the N\'{e}el phase while $K_{u(s)}^x =-\text{Im}K_{u(s)}^{+}$ is non-zero only in the IC phase.

\section{\label{sec:Conc}Conclusions}
We have studied the joint influence of longitudinal and transverse magnetic and electric fields on the ground state properties of the anisotropic Heisenberg spin-1/2 XXZ chain with large Ising anisotropy using DMRG method. We have obtained the ground state phase diagram in the presence of both the longitudinal and transverse magnetic fields. There are three different phases corresponding to a gapped N\'{e}el phase with AF order, gapped saturated phase and a critical incommensurate gapless phase when both longitudinal and transverse fields are present. In the presence of the transverse field, the nature of the behaviour of the uniform and staggered magnetizations near the critical fields changes from a cusp square-root singular behaviour for pure longitudinal field to a smooth behaviour. The external electric field does not lead to any new phase, however, the phase boundaries get modified. With increasing electric field, the AF N\'{e}el phase region reduces while the IC critical region grows. Electric field strengths comparable and larger than the transverse field tend to reduce the effect of the transverse field and restore the sharpness of the transitions near the phase boundaries. The electric field induces an electric polarization which takes its maximum value in the IC phase. 
We argue that the maximum value of the electric polarization occurs at field values where there is a crossover from an Ising like phase with dominant longitudinal spin correlations to an XY like phase with dominant transverse spin correlations. Interestingly, we show that even in the absence of an electric field, the transverse magnetic field induces a uniform and staggered nematic order in the fully saturated phase and a staggered vector chiral order in the N\'{e}el phase. Within the spin current mechanism, the induced vector chiral order can be identified with a staggered electric polarization in a direction parallel to the AF order. The presence of both uniform and staggered nematic order suggests that the transverse magnetic field may be used to induce a bond-order dimerization. 

Since our results indicate that the main effect of the electric field is to change the critical magnetic field strengths and the nature of the transitions between the different phases, it implies that both external magnetic fields and electric fields can be used to tune between different kinds of order and phases in the spin system. This can be tested experimentally by investigating the magnetization and electric polarization in  quasi-one dimensional Ising-like  magnetic systems ~\cite{shiba-ueda, kenzelmann, breunig2015, bera2015, *bera, nishiwaki2013, *nishiwaki2017} in the presence of applied magnetic and electric fields. Also, neutron scattering~\citep{oosawa, bera2015, *bera, matsuda2017} and electron spin resonance (ESR) experiments~\cite{kimura2007, zeisner} can be used to study the interplay of external electric and magnetic fields. Previous experimental studies showed that magnetic fields can induce an incommensurate order due to a crossover in quasi-1d antiferromagnets due to the crossover from an Ising like phase to a XY like phase ~\cite{kimura}. Our results suggest that an external electric field can be also used to tune such a crossover from an Ising like phase to a XY like phase and hence induce incommensurate order.

\acknowledgments{PD thanks BCUD, SP Pune University and SERB, DST India for financial support through research grant.}

\end{document}